\documentclass[12pt]{article}
\pdfoutput = 1
\begin{document}
\title
{Thermodynamics of black holes and the symmetric generalized uncertainty principle}

\author{
{\bf {\normalsize Abhijit Dutta}$^{a} $\thanks{dutta.abhijit87@gmail.com}}, 
{\bf {\normalsize Sunandan Gangopadhyay}$^{b,c}
$\thanks{sunandan.gangopadhyay@gmail.com, sunandan@iucaa.ernet.in}}\\
$^{a}${\normalsize Department of Physics, Adamas Institute of Technology, Barasat, Kolkata 700126, India}\\
$^{b}${\normalsize Department of Physics, West Bengal State University, Barasat, Kolkata 700126, India}\\
$^{c}${\normalsize Visiting Associate in Inter University Centre for Astronomy $\&$ Astrophysics,}\\
{\normalsize Pune, India}\\
[0.3cm]
}
\date{}

\maketitle

\begin{abstract}
{\noindent In this paper, we have investigated the thermodynamics of Schwarzschild black holes using the symmetric generalized uncertainty principle which contains correction terms involving momentum and position uncertainty. We obtain the mass-temperature relation
and the heat capacity of the black hole using which we compute the critical and remnant masses. The entropy is found to satisfy the area theorem upto leading order corrections from the symmetric generalized uncertainty principle.}
\end{abstract}

\noindent The understanding of the thermodynamic properties of black holes has been one of the most remarkable achievements in theoretical physics. Recently, the idea of a minimal length equal to the Planck length in various theories of quantum gravity \cite{str},\cite{nc} and quantum gravity phenomenology \cite{vag1}-\cite{nicolini} have led to a serious study of black hole thermodynamics \cite{hawk1}-\cite{baken} and its quantum corrected entropy \cite{adler}-\cite{rb}.  

In this paper we will study the thermodynamics properties of Schwarzschild black holes using the symmetric generalized uncertainty principle (SGUP) \cite{kim-son}
\begin{eqnarray}
\delta x\delta p&\geq&\frac{\hbar}{2}\left\{1+\frac{{\gamma}^2}{{L^2}}({\delta x})^2 
+\frac{\beta^2 l_{p}^2}{\hbar^2}(\delta p)^2\right\}
\label{sgup-1}
\end{eqnarray}
where $l_p$ is the Planck length ($\sim 10^{-35}m$), $\gamma$ and  $\beta$ are dimensionless constants and $L$ is also a minimal length
of the order of Planck length. 
We obtain the mass-temperature relationship from which we compute the heat capacity of the black hole. From this, we calculate the critical and remnant masses in terms of the Planck mass and the uncertainty constants for the black hole. We then compute the entropy keeping leading order corrections from the SGUP. The well known area theorem is recovered with corrections from the SGUP.

To start with we consider a Schwarzschild black hole of mass $M$. Near the event horizon of the black hole, the momentum uncertainty and the temperature for a massless elementary particle are related as \cite{adler}
\begin{eqnarray}
T=\frac{(\delta p) c}{k_B}
\label{mom_u}
\end{eqnarray}
where $c$ is the speed of light and $k_B$ is the Boltzmann constant. At thermodynamic equilibrium, the temperature of the black hole will be equal to that of the particle. 
Also, near the horizon of the Schwarzschild black hole, the position uncertainty of a particle can be expressed in terms of the Schwarchild radius \cite{adler},\cite{medved}
\begin{eqnarray}
\delta x=\epsilon r_{s}~;~
r_{s}=\frac{2GM}{c^2}
\label{position_u}
\end{eqnarray}
where $\epsilon$ is a calibration factor, $r_s$ is the radius of Schwarzschild black hole and $G$ is the Newton's universal gravitational constant.
\noindent To relate the temperature with the mass of the black hole, the GUP (\ref{sgup-1}) has to be saturated
\begin{eqnarray}
\delta x\delta p& = &\frac{\hbar}{2}\left\{1+\frac{{\gamma}^2}{{L^2}}({\delta x})^2 
+\frac{\beta^2 l_{p}^2}{\hbar^2}(\delta p)^2\right\}.
\label{sgup-2}
\end{eqnarray}
Using eqs.(\ref{mom_u}, \ref{position_u}), the above relation can be put in the following form
\begin{eqnarray}
M=\frac{M_{p}^2 c^2}{4\epsilon }\left\{\frac{1}{k_{B}T}+\frac{1}{k_{B}T}\frac{{\gamma}^2}{{L}^2}\left(\frac{2\epsilon M}{M_p}\right)^2\left(\frac{\hbar}{M_pc}\right)^2+ \frac{\beta^2}{(M_p c^2)^2}k_{B}T\right\}
\label{mass-temp1}
\end{eqnarray}
where the relations $\frac{c\hbar}{l_p}=M_{p}c^{2}$ and $M_{p}=\frac{c^2 l_p}{G}$ ($M_{p}$ being the Planck mass) has been used.
To fix $\epsilon$, we set the uncertainty constants $\gamma$ and $\beta$ to zero which gives
\begin{eqnarray}
M=\frac{M_{p}^2 c^2}{4\epsilon k_{B}T}~.
\label{mass-temp_no_grav}
\end{eqnarray}
Comparing this with the semi-classical Hawking temperature $T=\frac{M_{p}^{2}c^2}{8 \pi M k_{B}}$ \cite{hawk1}, \cite{hawk2}, yields the value of $\epsilon=2\pi$.
Hence eq.(\ref{mass-temp1}) can be written as
\begin{eqnarray}
M&=&\frac{M_{p}^2 c^2}{8\pi }\left\{\frac{1}{k_{B}T}+\frac{1}{k_{B}T}\frac{{\gamma}^2}{{L}^2}\left(\frac{8\pi M}{M_p}\right)^2\left(\frac{\hbar}{M_pc}\right)^2+ \frac{\beta^2}{(M_p c^2)^2}k_{B}T\right\}.
\label{mass-temp2}
\end{eqnarray}
Introducing the following relations 
\begin{eqnarray}
M'= \frac{8\pi M}{M_p}~~;~~T' = \frac{k_B}{M_p c^2}~~;~~L' = \frac{M_p c}{\hbar} L = \frac{L}{l_p}
\label{rel}
\end{eqnarray}
eq.(\ref{mass-temp2}) can be expressed as
\begin{eqnarray}
M'=\frac{1}{T'}+\frac{\gamma^2}{L'^2}\frac{M'^2}{T'}+\beta^2 T'.
\label{mass-temp_mod}
\end{eqnarray}
Now by definition, the heat capacity of the black hole is given by
\begin{eqnarray}
C = c^2\frac{dM}{dT}
\label{heat_cap_def}
\end{eqnarray}
which by using eq.(\ref{mass-temp_mod}) gives
\begin{eqnarray}
C = \frac{k_B}{8\pi}\left(\frac{2\beta^2 T'-M'}{T'-\frac{2\gamma^2}{L'^2}M'}\right).
\label{heat cap}
\end{eqnarray}
To get the remnant mass (where the radiation process stops), we set $C = 0$ and this leads to
\begin{eqnarray}
M_{rem} = \frac{\beta}{4\pi}M_p\sqrt{\frac{1}{1-\frac{4\beta^2 \gamma^2 \hbar^2}{L^2 M_p^2 c^4}}}~.
\label{remnant mass}
\end{eqnarray}
The condition that remnant mass is real leads to the following inequality 
\begin{eqnarray}
\frac{4\beta^2 \gamma^2 \hbar^2}{L^2 M_p^2 c^4} < 1.
\label{cond}
\end{eqnarray}
Now from eq.(\ref{mass-temp_mod}), we cn express the temperature in terms of the mass as 
\begin{eqnarray}
T' = \frac{M'-\sqrt{M'^2\left(1-\frac{4
\beta^2 \gamma^2}{L'^2}\right)-4\beta^2}}{2\beta^2}
\label{temp-mass}
\end{eqnarray}
where the negative sign before the square root has been taken to 
reproduce eq.(\ref{mass-temp_no_grav}) in the $\gamma, \beta\rightarrow0$ limit.
The above relation readily leads to the existence of a critical mass below which the temperature becomes a complex quantity 
\begin{eqnarray}
M_{cr} = \frac{\beta}{4\pi}M_p\sqrt{\frac{1}{1-\frac{4\beta^2 \gamma^2 \hbar^2}{L^2 M_p^2 c^4}}}~.
\label{critical mass}
\end{eqnarray}
Eqs.(\ref{remnant mass}) and (\ref{critical mass}) for critical and remnant masses are equal. and for both the masses condition in eq.(\ref{cond}) is applicable.
In the limit $\gamma\rightarrow0$, both the results reduces to the one found in \cite{abhi}. 

We now move to calculate the entropy which from the first law of black hole thermodynamics reads
\begin{eqnarray}
S=\int c^2 \frac{dM}{T} = \frac{k_B}{8\pi}\int\frac{dM'}{T'}
\label{entropy def}
\end{eqnarray}
where we have used eq.(\ref{rel}) to write the second equality.
Substituting eq.(\ref{temp-mass}) in eq.(\ref{entropy def}) and carrying out the integration expansion 
keeping terms up to leading order in $\gamma^2$ and $\beta^2$ yields
\begin{eqnarray}
 \frac{S}{k_B} &=& \frac{4\pi M^2}{M_p^2} -\frac{\beta^2}{8\pi}\ln\left({\frac{8\pi M}{M_p}}\right)-\frac{\gamma^2}{24 \pi L'^2}\left(\frac{8\pi M}{M_p}\right)^3 \nonumber\\
&=&\frac{S_{BH}}{k_B}-\frac{\beta^2}{16\pi}\ln\left(\frac{S_{BH}}{k_B}\right)-\frac{\beta^2}{16\pi}\ln(16\pi)-8\sqrt{\pi}\frac{\gamma^2}{3L'^2}\left(\frac{S_{BH}}{k_B}\right)^{\frac{3}{2}}
\label{entropy1}
\end{eqnarray}
where  $\frac{S_{BH}}{k_B} =\frac{4 \pi M^2}{M_p^2}$ is the semi-classical Bekenstein-Hawking entropy
for the Schwarzschild black hole. 
\noindent  In terms of the area of the horizon $A=4\pi r_{s}^2 =16\pi \frac{G^2 M^2}{c^4} = 4 l_p^2 \frac{S_{BH}}{k_B}$,
eq.(\ref{entropy1}) can be recast in the following form
\begin{eqnarray}
\frac{S}{k_B} = \frac{A}{4l_p^2}-\frac{\beta^2}{16\pi}\ln\left(\frac{A}{4l_p^2}\right)-\frac{\beta^2}{16\pi}\ln(16\pi)-\frac{8\sqrt{\pi}\gamma^2 l_p^2}{3L^2}\left(\frac{A}{4l_p^2}\right)^{\frac{3}{2}}
\end{eqnarray}
which is the famous area theorem with corrections from the SGUP.


We conclude by summarizing our findings. In this paper, we study the effect of the SGUP in the thermodynamics of Schwarzschild  black holes. We obtain the mass-temperature relation and the heat capacity of the black hole using which we compute the critical and remnant masses which are found to be equal and are consistent 
with our earlier findings \cite{abhi}, \cite{abhi1}. From the expression for the critical mass, we also obtain an inequality involving the uncertainty constants $\gamma$ and $\beta$. Finally, we compute the entropy and recover the area theorem with the SGUP corrections. We observe that the SGUP leads to a correction term of the form $A^{\frac{3}{2}}$.

\end{document}